# Formation and stability of a two-dimensional nickel silicide on Ni (111): an Auger, LEED, STM, and high-resolution photoemission Study

B. Lalmi[1*], C. Girardeaux[2], A. Portavoce[2], C. Ottaviani[3]
B. Aufray[4], J. Bernardini[2]

[1] - Synchrotron SOLEIL, L'orme des Merisiers, BP 48, 91192 Saint-Aubin, France
[2] - IM2NP (UMR 6242), CNRS, Aix-Marseille Université, Faculté des Sciences et Techniques, Campus de Saint-Jérôme F-13397 Marseille cedex 20, France
[3] - CNR-ISM, via Fosso del Cavaliere, 00133 Roma, Italy
[4] - CINaM-CNRS, Campus de Luminy Case 913 ; 13288, Marseille cedex 09, France

*Corresponding author: lalmi.boubekeur@gmail.com

## Abstract

Using low energy electron diffraction (LEED), Auger electron spectroscopy (AES), scanning tunnelling microscopy (STM) and high resolution photo-electron spectroscopy (HR-PES) techniques we have studied the annealing effect of one silicon monolayer deposited at room temperature onto a Ni (111) substrate.

The variations of the Si surface concentration, recorded by AES at 300°C and 400°C, show at the beginning a rapid Si decreasing followed by a slowing down up to a plateau equivalent to about 1/3 silicon monolayer.

STM images and LEED patterns, both recorded at room temperature just after annealing, reveal the formation of an ordered hexagonal superstructure $(\sqrt{3}\times\sqrt{3})R30°$-type. From these observations and from a quantitative analysis of HR-PES data, recorded before and after annealing, we propose that the $(\sqrt{3}\times\sqrt{3})R30°$ superstructure corresponds to a two dimensional (2D) $Ni_2Si$ surface silicide.

**Introduction**

In the last decade, the study of growth of ultrathin films has greatly progressed. Much of the effort on this matter has mainly been devoted to the growth of two dimensional (2D) layers with properties different from those of the bulk. Most of the works reported yet in the literature concerned the deposition of thin metallic films on semiconductor M/SC. For low coverage this leads to formation of ordered and well defined structures. Many different combinations of metals and semiconductors have been extensively studied [1-8].

The Ni-Si interface is a typical example of a system in which appear a number of surface phases during reactive diffusion. The initial stages of nucleation and growth of 2D and 3D phases at the Ni/Si interface have been extensively studied by structural or microscopy surface techniques, such as low-energy electron diffraction (LEED), low-energy electron microscopy (LEEM), scanning tunnelling microscopy (STM) [9-12]. Several 2D phases are reported to appear during the reaction of an ultrathin Ni film with a Si (111) substrate, $(\sqrt{3}\times\sqrt{3})R30°$, $(\sqrt{19}\times\sqrt{19})R23.4°$ and $(1\times1)-RC$ phases [13-17]. The last two phases are the most likely intermediate steps to epitaxial growth of 3D NiSi$_2$ onto Si (111) [18, 19, 20]. On Si (001), depending on the Ni coverage and thermal annealing conditions, several nickel-silicide compounds can be formed [22]. Yoshimura et al [21] have shown by STM that adsorbed Ni atoms immediately react with Si substrate forming a $(2\times1)$ 2D alloy. This reconstruction is induced by the dimerization of Si dangling bonds under the top Ni layer [22, 23, 24]. For Ni coverage over to 0.7ML, NiSi$_2$ islands coexist with the $(2\times1)$ structure.

For Si (111) and Si (100), when the Ni coverage exceeds one monolayer the surface morphology is always a miscellaneous of 3D islands and 2D structures [16, 20, 22].

The bulk phase diagram of the Ni-Si system is asymmetric; indeed, whereas the solubility limit of Ni in Si is negligible, the solubility of Si in Ni is about 10% at 700°C [25]. That means a possible Si diffusion in the bulk of Ni, which is not the case in the reverse system. In spite of many efforts to understand the structure and reaction kinetics for the Ni–Si system, there are only a few reports discussing the "reverse system", i.e. reaction of ultrathin silicon films onto nickel surfaces (Si/Ni). Yet, understanding of the non-equivalence between the sequences of deposition Ni/Si and Si/Ni is of great interest for the comprehension of the initial stages of Schottky barrier formation.

In the present study, we are interested in the early stages of Si adsorption and reaction onto Ni (111) surface and, in particular, in formation of possible ordered overlayer and/or interface structure which may be induced by thermal annealing. In detail, we examine, using four surface sensitive techniques, AES, LEED, HR-PES, and STM, the precise composition, structure and kinetics properties of the nickel-silicon interface formed, when one silicon monolayer is deposited onto Ni (111).

**Experimental details**

All experiments were carried out in ultrahigh vacuum (UHV) conditions with a base pressure below $2 \times 10^{-10}$ Torr. They were performed in two different experimental equipments. AES, LEED and STM analysis have been performed at CINaM, Marseille. Auger spectra were obtained in the derivative mode, and the data collected with a computer system allowing an easy measurement of the peak-to-peak height of the Auger signal of elements close to the surface versus annealing time (or deposition time). The LEED optics placed in the same chamber was used to follow the surface structure evolutions. The STM (Omicron STM1) microscope, which is in the main chamber, was used at room temperature after annealing.

HR-PES experiments were performed at the "VUV" beam line of the ELETTRA synchrotron radiation facility in Trieste, Italy. Electron distribution curves were recorded with a hemispherical energy analyser. To study the $Si_{2p}$ core levels, the photon energy was set to E= 177.7 eV. Before any experiment, whatever the technique used, the Ni (111) surface was cleaned by cycles of $Ar^+$ ion bombardment ($5 \times 10^{-5}$ Torr at 500 eV) followed by annealing at 750°C. The sample temperature was evaluated from a thermocouple spot-welded on the sample holder very close to the crystal. The sputtering-annealing cycles were performed until a sharp $(1\times1)$ LEED pattern was observed. Silicon deposition was carried out in the main chamber by thermal evaporation from a silicon wafer heated by Joule effect. The Si coverage has been estimated from the growth kinetics previously recorded at RT by AES on a Cu substrate [26, 27]. On this substrate, the growth is close to a layer-by-layer mode, at least up to five ML. On the experimental curves, as well as on the simulated curves, the first break, which corresponds to the completion of the first ML, appears when the intensity of the Cu (60 eV) Auger signal is attenuated of about 60%. On the simulation of the experimental curve, the Si ML was assumed to be a dense plane with a depth of 0.235 nm. In the present study, we supposed that the first Si ML is obtained when the same attenuation is reached in the Ni (61 eV) Auger signal.

**Results and discussions**

Isochronal dissolution kinetics of one silicon monolayer (1 Si ML) has been recorded in the temperature range [50–650°C] with an annealing rate of about 1.5°C/min. During the temperature rise, the evolution of both Ni (61 eV peak) and Si (92 eV peak) Auger signals were monitored by AES measurements. Fig.1 shows the variation of the Si (92eV) to Ni (61eV) Auger signals intensity ratio versus temperature. Three domains can be delimited on this curve. From the room temperature (RT) up to about 130°C (domain I), one observes a fast decrease of the Auger signals $I_{Si}/I_{Ni}$ ratio. This first decrease is followed by a slower decreasing (domain II) in the temperature range [130°C–440°C]. In the third part (domain III), the curve reaches a plateau corresponding to a $I_{Si}/I_{Ni}$ ratio close to 0.25.

During the temperature rise, the evolution of the surface structure was also followed by LEED. Fig.2-a shows a typical sharp $(1\times1)$ LEED pattern ($E$=64 eV) recorded on a clean Ni (111) face after the surface preparation. After deposition of 1 Si ML, the $(1\times1)$ LEED pattern is still there but the spots are more fuzzy and the background more intense which is characteristics of an unstructured deposit (Fig.2-b); From about 130°C, a $(\sqrt{3}\times\sqrt{3})R30°$ superstructure starts to appear (Fig 2c) to become more and more intense and sharper as the temperature increases (Fig 2d). Note that this superstructure is observed up to 650°C.

It is generally difficult to interpret quantitatively an isochronal dissolution kinetics because many parameters evolve with time and temperature (bulk diffusion coefficient $D_b$, limit of the solubility $C_{b\ limit}$,…), mixing kinetics to thermodynamics phenomena. However, it often shows clearly the global behaviour of the system, high lighting the temperature domains where interesting phenomena occur. Nevertheless the isochronal kinetics presented in Fig.1 can be roughly understood as follow. Because the fast decrease of the $I_{Si}/I_{Ni}$ ratio observed in the first domain (T<130°C) cannot be linked to silicon bulk dissolution (the temperature is too low) it is more likely the signature of a reaction/reorganisation of the silicon ML with the very first Ni atomic layers. Such a reaction/reorganisation characterized by a fast decrease in the ratio of Auger signals can correspond to (1) the formation of a thin 3D surface alloy (silicide) covering the entire surface (the superstructure observed by LEED at the end of the first domain would be the signature of a perfect epitaxy between this silicide and the Ni(111) substrate) (2) islanding of the unreacted Si deposit in equilibrium with a 2D superficial compound which is not very likely because the Ni-Si system presents a strong tendency to form alloy and not to phase separation or to (3) formation of 3D silicide clusters which do not cover the entire surface in equilibrium with a superficial 2D compound (since a superstructure is observed at the end of the domain). Concerning the domain II, where a slowdown is observed in the decrease of the Auger intensity ratio, two scenarios can be foreseen. The first one, based on the idea that no islanding appears at the end of the first domain (hypothesis (1) exposed previously), is the dissolution of a thin 3D surface alloy (silicide) covering the entire surface. The decrease observed in the domain II corresponds then to a continuous dissolution of this thin 3D alloy (silicide) up to the formation of a 2D surface alloy (domain III) more stable than the 3D one. The second scenario would correspond to a decrease of the size of the 3D clusters up to complete dissolution (via silicon dissolution in the bulk of Nickel) in equilibrium with the 2D compound forming the superstructure observed by LEED. In any case, the kinetics blocking observed in the first part of the domain III can be explained by a stronger stability of the 2D compound forming the $(\sqrt{3}\times\sqrt{3})R30°$ superstructure linked to the strong Si surface segregation tendency due to its lower surface energy in comparison with the nickel one [28]. Results of Photo Electron Spectroscopy presented later will allow specifying the scenario.

Using the same technique (AES-LEED), isothermal dissolution kinetics have been recorded at two temperatures (~300°C and ~400°C). The variation of Auger peak-to-peak intensity Si (92eV) to Ni (61eV) ratio versus time for each temperature is displayed in Fig. 3a and 3b.

The LEED observations carried out at room temperature at the end of each kinetics exhibit the same sharp and well defined $(\sqrt{3}\times\sqrt{3})R30°$ superstructure as observed at the end of the isochronal kinetics.

On the kinetics recorded at 300 °C, one can observe again three domains: (i) the rapid decreasing of the $I_{Si}/I_{Ni}$ ratio at the beginning of annealing, probably due to the reaction/reorganisation previously detailed of the silicon ML with the very first Ni atomic layers, then (ii), a slow continuous dissolution of a 3D compound (or 3D clusters) and (iii) the kinetics blocking on the $(\sqrt{3}\times\sqrt{3})R30°$ superstructure. On the kinetics recorded at 400°C, domain I and domain II are probably too close (mixed) to be observed. It is interesting to note that at the end of both isothermal kinetics and at the end of isochronal kinetics, one observes the same superstructure and the same surface concentration ($I_{Si}/I_{Ni}$ ~ 0.25). If one accept that this ratio is mainly characteristic of Si atoms staying on surface, it corresponds to about 1/3 of Si ML (2/3 have been dissolved in the Nickel volume).

From isothermal dissolution kinetics it is possible to derive the order of magnitude of the bulk diffusion coefficients involved during the first part of the dissolution process. This estimation can be done easily from the time that the system uses to dissolve in the bulk around 2/3 of the Si ML (~80 min at 300°C and ~30 min at 400°C) with the assumptions that there is no evaporation as well as no islands formation of silicon.

Using the relation [29, 30, 31]:

$$\Delta C_s(t) = 2 C_{b\,limit} \sqrt{\frac{D_b t}{\pi}}$$

where $D_b$ is the bulk diffusion coefficient, $t$ the time to dissolve 2/3 of the Si ML ($\Delta C_s$ the variation of Si amount after a time t) and $C_{b\,limit}$ the limit of the solubility of Si in Ni (given by the bulk phase diagram), we have derived both bulk diffusion coefficients $D_b(300°C) \sim 6 \times 10^{-18}$ cm$^2$.s$^{-1}$ and $D_b(400°C) \sim 1 \times 10^{-17}$ cm$^2$.s$^{-1}$. These values are very high in comparison with the values expected from extrapolation of high temperature measurement which are respectively $4 \times 10^{-24}$ cm$^2$.s$^{-1}$ and $1 \times 10^{-20}$ cm$^2$.s$^{-1}$ [32].

From this simple evaluation of the bulk diffusion coefficients we can deduce that the first parts of the kinetics are not only linked to a bulk dissolution process but also to a reaction/reorganisation of Si atoms with the first Ni layers.

Note that similar kinetics behaviours have been also observed on equivalent bi-metallic systems i.e. with a tendency to order and to a strong surface segregation of the deposited element [33, 34, 35].

This reorganisation/reaction phenomenon of Si at the Ni (111), followed by a partial dissolution process (up to the superstructure) seems to deeply mark the topography of the surface. Indeed, filled state STM images (Fig.4-a) recorded after the dissolution process at 400°C (annealing time 45min) and the formation of the $(\sqrt{3} \times \sqrt{3})R30°$ superstructure (180×180nm$^2$) appear very corrugated with many holes (small and large). The depth profile reported in figure 4-b gives the average depth of these holes, which is approximately equal to 0.2 nm. These holes, not present on the clean Ni (111) surface (not shown here), are likely the mark of the first steps of the Si deposition process followed by dissolution of the 3D silicide islands.

A magnification taken on a flat part of Figure 4 shows the $(\sqrt{3}\times\sqrt{3})R30°$ superstructure atomically resolved (filled state STM image reported on Figure 5).

On this STM image the average distance separating two successive rows in the periodic arrangement is about 0.43 nm, which corresponds to $\sqrt{3}\ a_{Ni}$ (with $a_{Ni}$= 0.249 nm the distance between firsts nearest neighbours on the Ni (111) surface). On these rows, the corrugation amplitude is close to 0.02±0.01nm as shown on the line scan reported on figure 5. This weak corrugation suggests that the atoms inducing the $(\sqrt{3}\times\sqrt{3})R30°$ superstructure, are inserted in the plane of the topmost layer forming a 2D surface alloy (2D silicide) as it is generally observed on similar systems [36, 37].

Only from the STM image, it is impossible to say if the 2D silicide is $Ni_2Si$ or $NiSi_2$. However, dissolutions kinetics (isochronal and isothermal) monitored by AES (Fig.1, Fig.3) show that the sharper $(\sqrt{3}\times\sqrt{3})R30°$ LEED pattern is obtained after dissolution of about 2/3 of Si ML, which is in favour of rich Ni compound, i.e. $Ni_2Si$.

Finally, we have performed at RT high-resolution photoelectron spectroscopy (HR-PES) measurements of the $Si_{2p}$ core-levels. Fig.6a and 6b present two photoemission spectra of the $Si_{2p}$ core-levels, recorded respectively just after deposition at RT of one Si ML onto a clean Ni(111) substrate and after annealing at 400°C during 10 minutes (with the $(\sqrt{3}\times\sqrt{3})R30°$ superstructure checked by LEED). Both spectra are obtained with the same energy of photons ($E$= 177.7 eV) and the same geometry (emission angle 45°, acceptance angle 16°). The binding energies are referenced with regard to the Fermi level and the peaks fitted with a Donia-Sunjic [38] peak line shape and a Shirley type [39] background. The best fit is obtained with the following parameters: (i) a Gaussian and a Lorentzian with a Full Width at Half Maximum (*FWHM*) respectively of 0.15 eV and 0.08 eV, a spin-orbit splitting of 0.60 eV, a branching ratio of 0.5 and an asymmetry parameter of 0.05.

With these parameters, the peak after deposition of one Si ML appears to be composed of five components: an intense component $S_1$ (99.06 eV), and four other components $S_2$ (99.13 eV), $S_3$ (99.23 eV), $S_4$ (98.77 eV) and $S_5$ (99.45 eV) showing the co-existence of different chemical environments of Si atoms. The $S_1$, $S_2$, $S_3$, $S_4$, $S_5$, components are pointed at the bottom of every spectrum on Fig. 6a. For comparison, the binding energy position of the $Si_{2p3/2}$ peak measured on a Si (111) (bulk peak) in our experimental configuration is also pointed by a full line ($E_b$= 99.3 eV). The $S_5$ component ($El$ = 99.45 eV) close to that of the Si bulk (99.3 eV), can be attributed to the Si atoms rather bounded to other Si atoms and at the topmost surface layer. The ($S_1$, $S_2$, $S_3$, $S_4$) components are more or less strongly shifted in energy toward the lower binding energy. This reflects an electron transfer between silicon and nickel and suggests that these Si atoms are in strong interaction with the nickel atoms; each component could be a signature of the number of Ni atoms, and more a silicon atom is surrounded by nickel atoms, and more its core level will be shifted toward lower binding energy. This also shows that the first Si atoms instantaneously react with the Ni substrate to form intermetallic compounds unstructured or ordered at up to a short distance, and thus not observable by LEED.

Using the same set of parameters, the $Si_{2p}$ core levels recorded after annealing at 400°C can be fitted with only one doublet as shown on Fig.6-b. Existence of only one component indicates that the Si atoms are in a single chemical environment, i.e. every Si atom in the surface layer is surrounded by a same number of Ni atoms forming the ordered $(\sqrt{3}\times\sqrt{3})R30°$ surface alloy observed by LEED. On the topmost surface layer, this corresponds to the 2D silicide $Ni_2Si$. Note that this component is already present at RT (and is the most intense) just after silicon deposition (Figure 5a). This shows the natural tendency of the system to form locally a 2D silicide with a composition close to $Ni_2Si$. Note also, that this $Si_{2p}$ core level spectrum presents a large asymmetry parameter (0.05). This asymmetry is the signature of a high metallic character of silicon on nickel since it is associated to a high density of states at the Fermi level. This metallic character of the silicon atoms can explain the corrugation observed on the STM images (Fig. 5), showing mainly the silicon atoms forming the $(\sqrt{3}\times\sqrt{3})R30°$ superstructure.

One can note that the total integrated intensity, the $Si_{2p}$ peak decreases of about 2/3 after the annealing treatment (the total area of the Si2p components decreases from 3.8 arbitrary unit - a.u. -, just after deposition, to 1.15 a.u after annealing), which is in very good agreement with the previous AES data presented in the first section and confirms the chemical composition of the 2D silicide (Ni2Si).

Finally, from all results (Auger, LEED, STM and PES), we can propose the following scheme (fig. 7):

I) Most of the silicon atoms (for a monolayer deposit) react instantaneously, during deposition with the nickel substrate, to form unstructured intermetallic compounds or ordered compounds at short distance as schematically shown in Fig. 7b and 7c (PES + LEED).

II) In the domain I (observed by AES during isochronal and isothermal dissolution), 3D silicide islands are formed in equilibrium onto a 2D surface silicide (Fig 7d).

III) In the domain II, the 3D silicide islands dissolve (diffusion of Si atoms at the interface between the islands and the substrate) until a 2D silicide. The surface roughness observed by STM (about 0.2 nm, Fig. 4) could be the signature of insertion of Si atoms in the topmost Ni surface layer (during the first step of deposition as schematically shown in Fig. 7b) followed by the dissolution process of the 3D islands (Fig. 7e).

IV) In domain III, the dissolution kinetics is blocked on a 2D silicide as shown by AES, PES, LEED and STM (Fig. 7f).

**Conclusions**

We have studied by AES-LEED, HR-PES and STM the temperature effect on 1 ML of Si deposited onto Ni (111). The Si deposition at RT leads to the formation of a disordered 2D and/or 3D intermetallic compound. The main effect of the isothermal or isochronal annealing is to dissolve the excess of Si in the bulk and to form a stable and well ordered 2D $Ni_2Si$ intermetallic compound giving rise to a $(\sqrt{3}\times\sqrt{3})R30°$ superstructure. In this surface alloy, Si is found to be inserted in the topmost surface layer with a strong metallic character. This surface alloy shows a stability with the temperature at least up to 400°C.

**Figure captions**

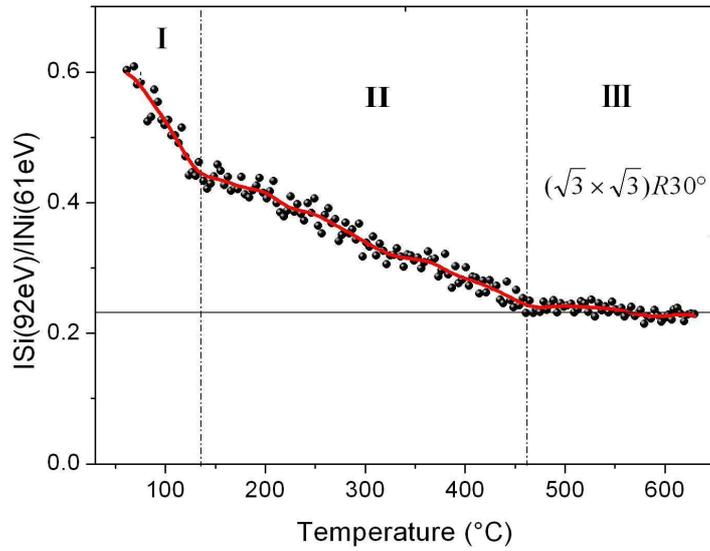

**Fig.1:** Variation of the Auger peak-to-peak intensity ratio ($I_{Si}I_{Ni}$) versus temperature during isochronal dissolution of 1 Si ML deposited onto Ni(111).

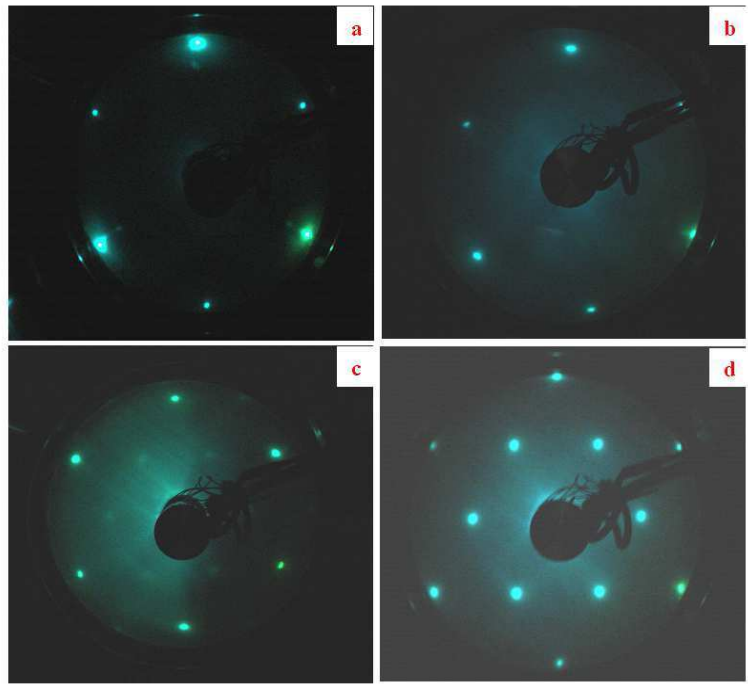

**Fig.2 :** LEED patterns (*E=64eV*) of a Ni(111) surface (a) on bare surface, (b) after deposition of one Si monolayer, (c) after annealing at 130°C, (d) after annealing at 400°C.

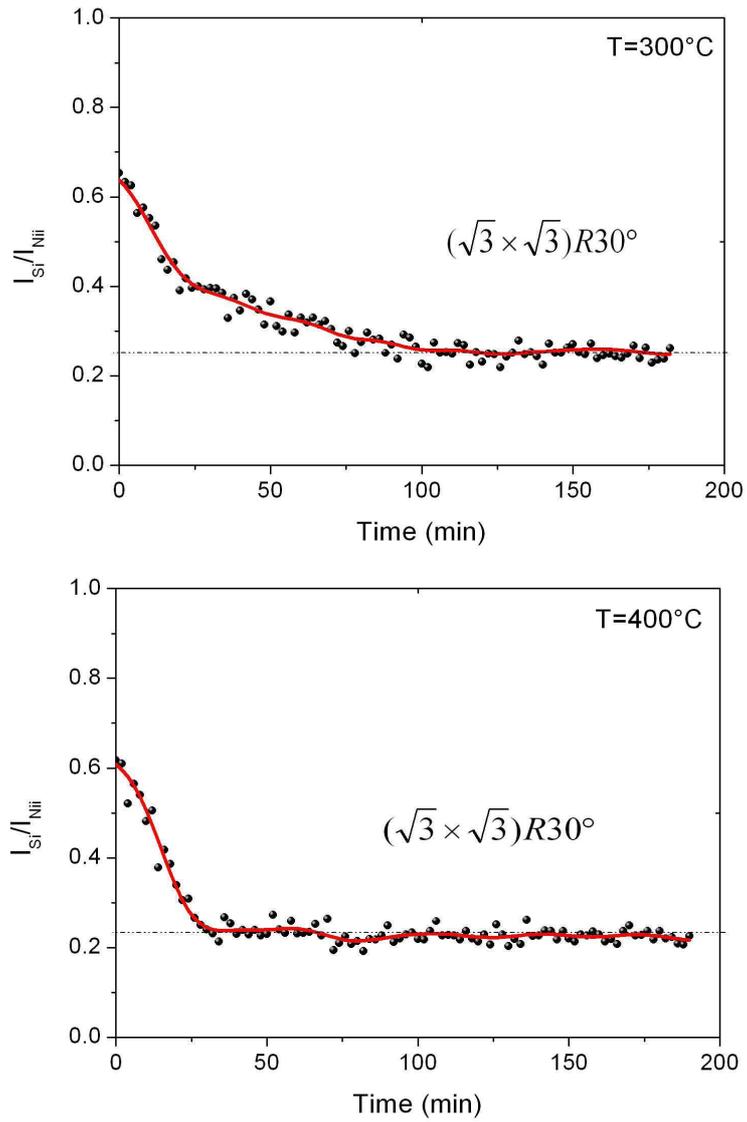

**Fig.3 :** Variation of the Auger peak-to-peak intensity ratio ($I_{Si}/I_{Ni}$) versus time during isothermal dissolutions of 1ML Si/Ni(111), a) at 300°C, b) at 400°C.

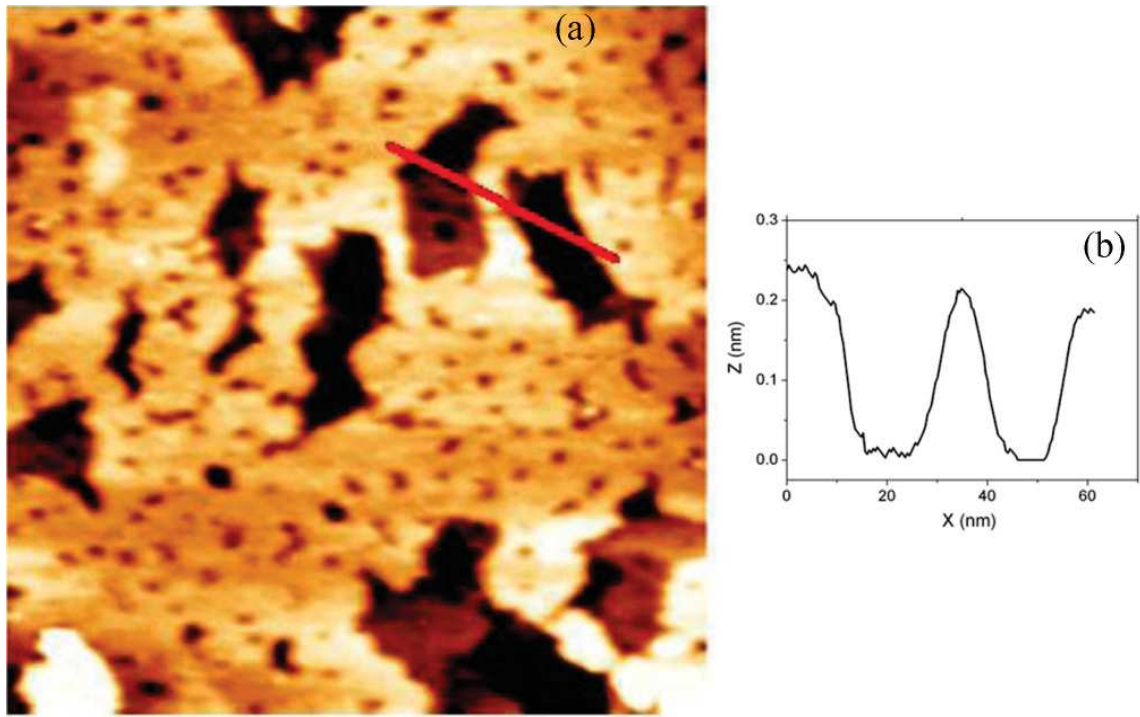

**Fig.4:** a- STM image of the Ni (111) surface covered by 1 Si ML after annealing at 400°C (Imaging conditions: 0.9V sample bias and 1.2nA tunnel current), b- line scan showing the depth holes.

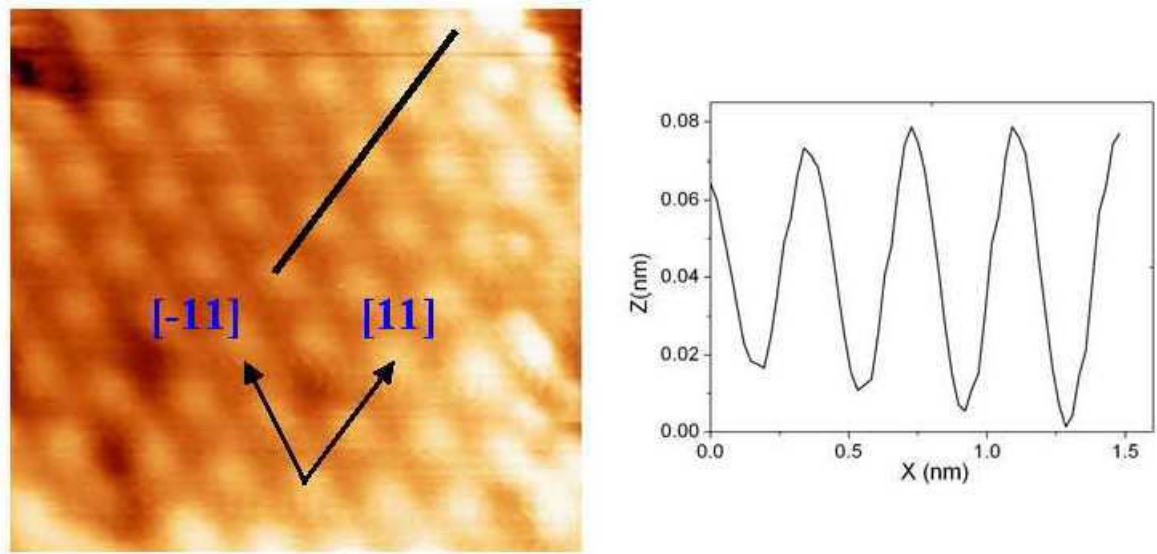

**Fig.5:** A typical filled states STM image of the Ni (111) surface showing the $(\sqrt{3} \times \sqrt{3})R30°$ superstructure obtained after annealing at 400°C for 10 min of 1 Si ML (0.5V ; 1.4nA).

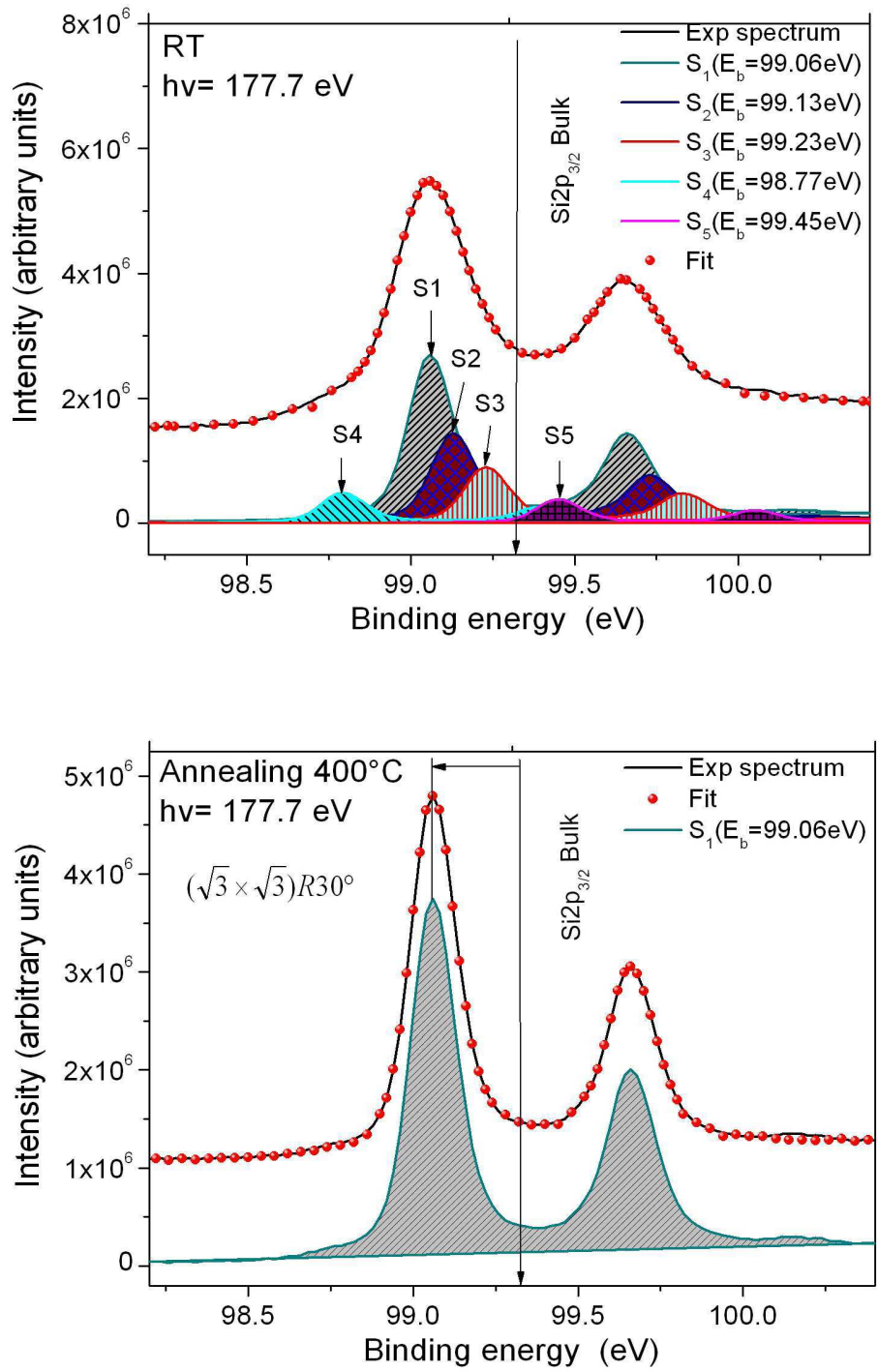

**Fig.6:** Si$_{2p}$ core levels a) after deposition at RT of one monolayer of Si onto Ni(111) b) after annealing at 400°C (45 min).

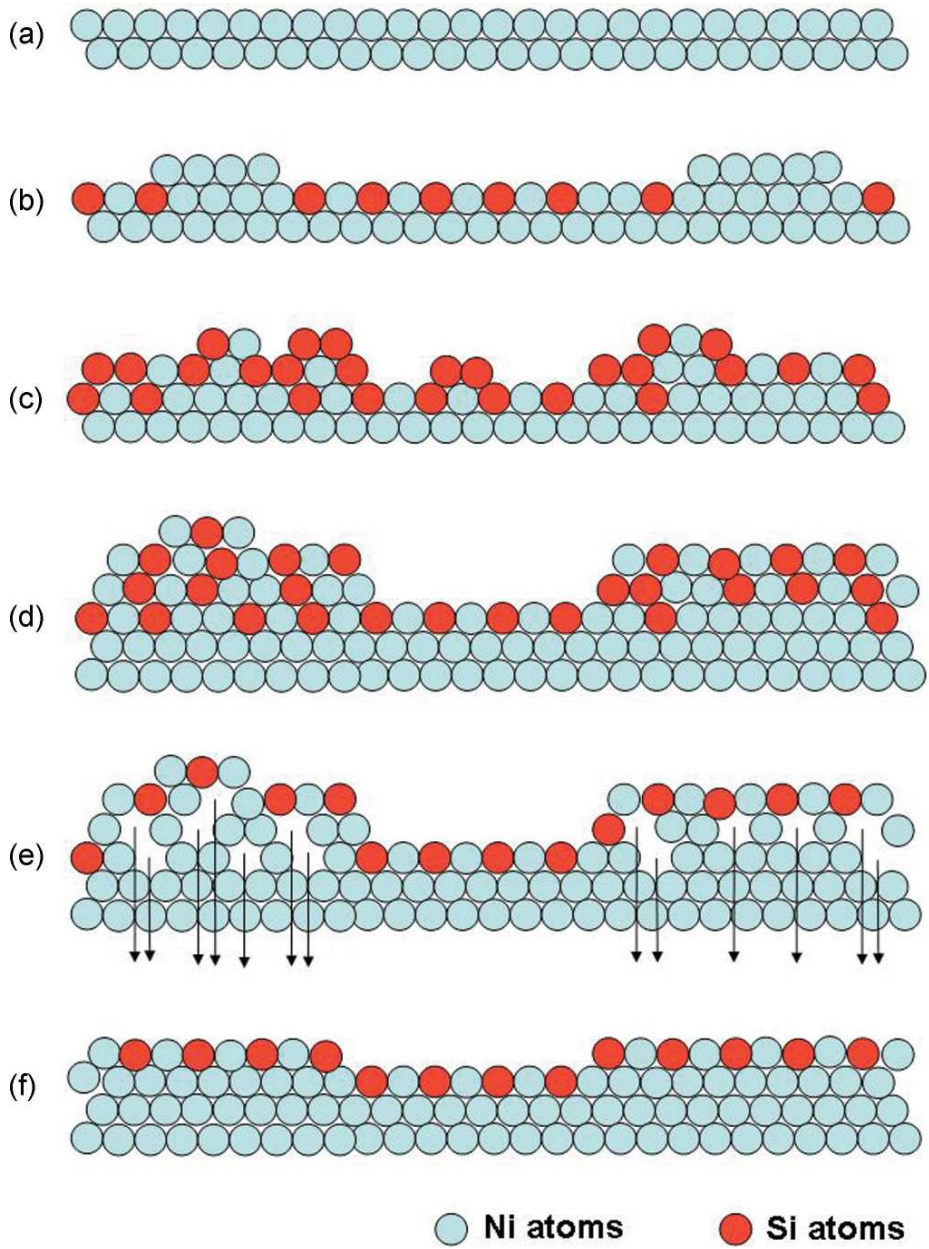

**Fig.7 :** Side view schematic atomic model suggesting the scenario observed during deposition and dissolution of about one Si Ml on Ni(111).
a- Large terrace of bare Ni(111)
b- First step of the Si deposition (formation of Ni terraces due to Si atoms insertion process)
c- After deposition of about 1 Si ML
d- Domain I observed by Auger during isochronal and isothermal annealing: formation of 3Dsilicide islands
e- Domain II : dissolution in the bulk of the Si atoms not located in the topmost surface layer due to the segregation phenomenon (i.e. dissolution of the 3D silicide islands)
f- Domain III: blockage of the dissolution process on a 2D silicide resulting in a surface roughness.